\documentclass{nature}
\usepackage{graphicx}
\usepackage{amsmath}
\usepackage{epstopdf}
\usepackage{float}
\usepackage{xcolor}
\usepackage{ccaption}
\usepackage{url}
\usepackage{textcomp}
\usepackage{gensymb}
\usepackage{hyperref}
\hypersetup{colorlinks=true,linkcolor=red,citecolor=blue}

\usepackage{amssymb}

\makeatletter
\let\saved@includegraphics\includegraphics
\AtBeginDocument{\let\includegraphics\saved@includegraphics}
\renewenvironment*{figure}{\@float{figure}}{\end@float}
\makeatother

\newcommand{\beqa}{\begin{eqnarray}}
\newcommand{\eeqa}{\end{eqnarray}}

\title{A single-atom level mechano-optical transducer for ultrasensitive force sensing}

\author{Yang Liu$^{1,2,*}$, Pengfei Lu$^{1,*}$, Xinxin Rao$^{1}$, Hao Wu$^{1}$, Kunxu Wang$^{1}$, Qifeng Lao$^1$, Ji Bian$^1$, Feng Zhu$^{1,2}$,  Le Luo$^{1,2,3}$}

\begin{document}
\maketitle

\begin{affiliations}
\item School of Physics and Astronomy, Sun Yat-sen University, Zhuhai 519082, China
\item Center of Quantum Information Technology, Shenzhen Research Institute of Sun Yat-sen University, Nanshan Shenzhen 518087, China
\item International Quantum Academy (SIQA), and Shenzhen Branch, Hefei National Laboratory, Futian District, China
\item $^\ast$These authors contributed equally to this work
\item E-mail:luole5@mail.sysu.edu.cn
\end{affiliations}


\begin{abstract}
Using light as a probe to detect a mechanical motion is one of the most successful experimental
approaches in physics. The history of mechanical sensing based on the reflection, refraction and
scattering of light dates back to the 16th century, where in the Cavendish experiment, the angle of
rotation induced by the gravitational force is measured by the deflection of a light beam reflected
from a mirror attached to the suspension. In modern science, mechano-optical transducers are such
devices that could detect, measure and convert a force or displacement signal to an optical one,
and are widely used for force detection. Especially, ultraweak force sensor with ultrahigh spatial
resolution is highly demanded for detecting force anomaly in surface science, biomolecule imaging,
and atomtronics. Here we show a novel scheme using a single trapped ion as a mechano-optical
transduction. This method utilizes the force-induced micromotion, converting the micromotion to a
time-resolved fluorescence signal, in which the ion's excess micromotion coupled to the Doppler
shift of the scattered photons. We demonstrate the measurement sensitivity about 600 $\textrm{zN}/\sqrt{\textrm{Hz}}$
(1 $\textrm{zN} =10^{-21}$N). By alternating the detection laser beam in all three dimensions, the
amplitude and the direction of a vector force can be precisely determined, constituting a 3D force
sensor. This mechano-optical transducer provides high sensitivity with spatial resolution in
single-atom level, enabling the applications in material industry and the search for possible
exotic spin-dependent interactions that beyond the standard model.
\end{abstract}
\maketitle

Force sensors with high sensitivity have rich applications in fundamental science and engineering.
The ability to measure extremely small forces plays a vital role in absolute gravimetry and
inertial navigation~\cite{niebauer1995new,bidel2013compact,battelier2016development}, determination
of gravitational constant $G$ \cite{cavendish1798xxi},  gravitational wave detection
~\cite{abbott2017gw170817}, force microscopy~\cite{binnig1986atomic}, and spin-resonance
imaging~\cite{rugar2004single} et.al. Dating back in history, since Henry Cavendish's first
measurement of the specific gravity of the Earth using torsion balance in 1797, a variety of
methods of measuring force have been developed. Among those methods, translating a mechanical
signal to an optical signal have been proved to be one of the most sensitive ones, especially
recent techniques of using both lasers and quantum sensors for mechano-optical systems, such as
micro- and nano-mechanical oscillator~\cite{stowe1997attonewton,mamin2001sub}, optically levitated
microspheres~\cite{geraci2010short}, and trapped ions~\cite{blatt2008entangled}. All the systems
have been applied for sensing very weak electric-magnetic, optical, gravitational and exotic forces. Among these
mechano-optical systems, trapped ions enable exquisite control of both spin and motional degrees of
freedom~\cite{maiwald2009stylus,biercuk2010ultrasensitive,knunz2010injection,blums2018single}, and
thus allowing quantum-enhanced measurement, therefore it is the most capable platform for
detecting ultraweak forces with ultrahigh spatial resolution.

Currently, there are three typical force-sensing scenarios developed with trapped ions. The first
one is to image the ion's displacement induced by the force, such as the ultrafast single-photon
counting ~\cite{knunz2010injection} and super-resolution
imaging~\cite{wong2016high,blums2018single}. The second one utilizes the Doppler effect, where the
force to be measured affects the frequency of the fluorescence photon,including phase-coherent
Doppler velocimetry~\cite{biercuk2010ultrasensitive}, quantum lock-in detection of the
fluorescence~\cite{shaniv2017quantum,gilmore2017amplitude,gilmore2021quantum}. The third method
employs quantum-enhanced techniques to measure the change of the spin state induced by the force.
The experiments along this route include measurements of the Fock states overlaps
~\cite{wolf2019motional}, amplification of coherent displacements by squeezing
~\cite{burd2019quantum}, many-body spin echo by entangling the motion mode and the collective
spin\cite{gilmore2017amplitude,gilmore2021quantum}. Based on these techniques, trapped-ion based
force sensor has been proposed for various applications. In one hand, it has been suggested for a
sensitivity in the yocto-Newton regime, a regime where quantum gravity can be experimentally
tested~\cite{kafri2014classical,albrecht2014testing}. On the other hand, instead of perusing the
extreme sensitivity, trapped ions have been proposed for 3D sensing of the force anomaly, where the
information of both direction and frequency of the force are lacking before the data acquisition.
Therefore, most of the schemes mentioned above are not suitable, since they are designed to measure
the forces that are periodically driven.

Here we present a new schemes based on the mechano-optical transduction without requiring
modulating the force to be measured, in which the exerted forces acted on a trapped ion can be
measured by the fluorescence signal related to the ion's micromotion. With the achieved sensitivity
of 229.17 $\textrm{zN}/\sqrt{\textrm{Hz}}$, our scheme could enable 3D sensing of the force anomaly. It is noted that
our method is very different with the force detection using Doppler velocimetry demonstrated in a
Penning trap~\cite{biercuk2010ultrasensitive}, where the ion's motion is modulated by AC electric
force with the axial center-of-mass mode frequency. The method presented in this paper does not
require any external modulation of the force. Instead, the force to be measured results in excess
micromotion that intrinsically has the same frequency of the driving frequency of the RF field used
for trapping, so called the RF-photon correlation~\cite{berkeland1998minimization}. With this
technique, by scanning the photon arrival time, phase information of fluorescence is natually
captured, thus providing a lock-in-like measurement of the force. Different with the standard
lock-in method, in which the signal to be measured must be modulated and the relative phase between
the modulation and demomodulation signal are fixed, the RF-photon correlation method can measure a
unknown force without modulation. This advantage provide a feasible scheme for sensing of the force
anomaly, where the frequency spectrum of the force is usually unknow before the measurement.

\begin{figure}
\centering
\includegraphics[width=0.8\textwidth]{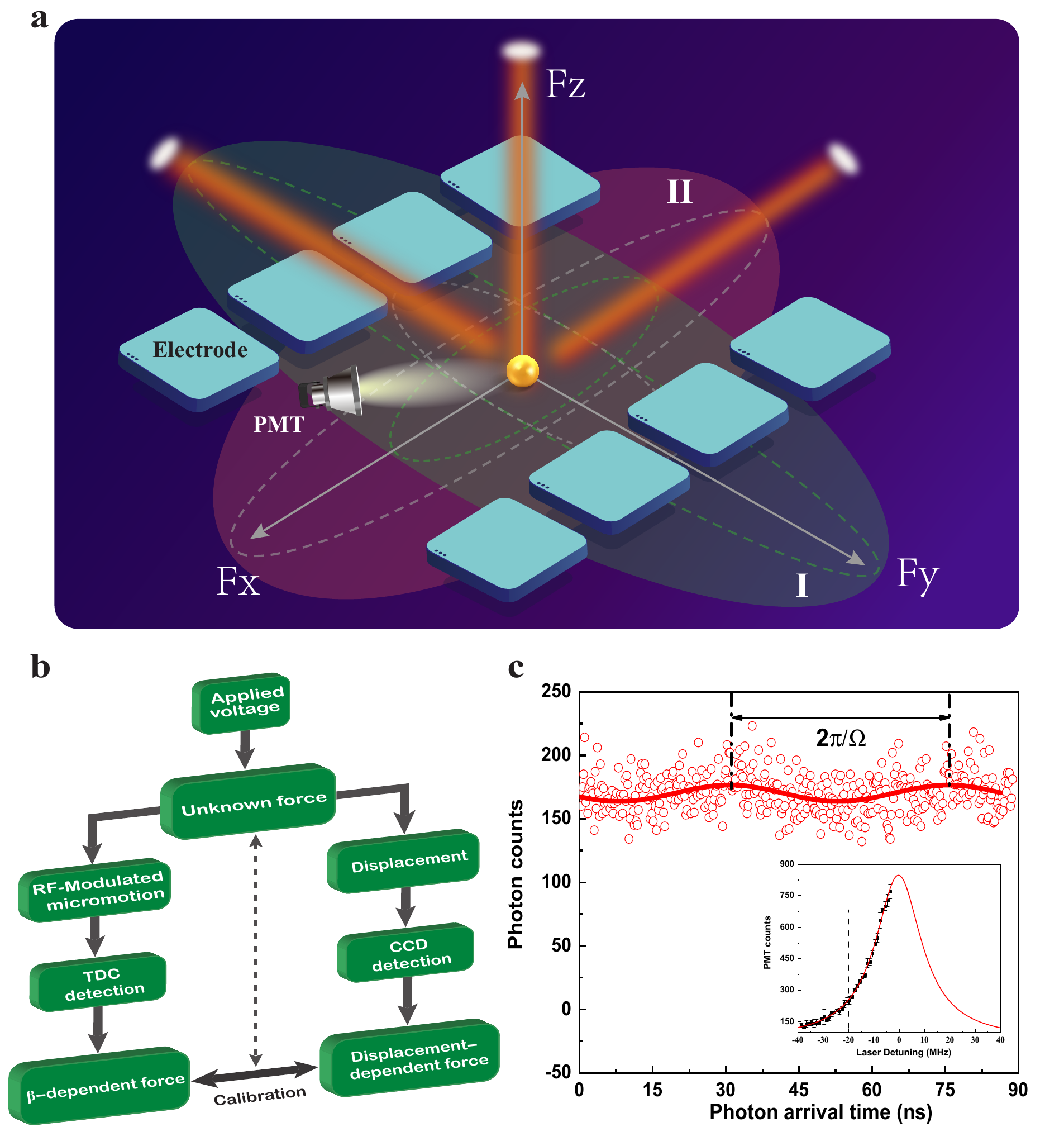}
\caption{Force sensor based on a trapped ion's excess micromotion. a, The basic schematics of force measurement using mechano-optical transducer. With detection laser along trap axis (or either of the radial principle axis), the force-induced RF-modulated fluorescence is collected by a photomultiplier tube (PMT), by which the three-dimensional geometry of the applied force, e.g. geometry $I$ (ellipsoid in light green) or $II$ (ellipsoid in light purple) can be found. b, The experimental flow chart of the force sensing using the mechano-optical transduction scheme. c, The ion's RF-photon correlated fluorescence (red circles), where the histogram of photon arrival times relative to start pulses generated synchronously with the RF driving field is shown. Photon arrivals are bunched with a periodicity dictated by the driven RF frequency. The solid red line is an theoretical fit to the data used to obtain the micromotion index $\beta$. The inset shows the resonance profile of the electronic transition  $^2S_{1/2}\leftrightarrow$ $^2P_{1/2}$. The black square and red line corresponds to the measurement and theoretical fit. Here, the detuning of the detection laser is chosen as $\Delta = -2\pi \times 20$ MHz $\approx-\Gamma$ marked by the vertical dashed line, in order to obtain a relatively large photon-correlation signal $\frac{\Delta}{S_0}$ \cite{keller2015precise}. The power of the detection laser throughout the experiment is kept in the low intensity limit, i.e. the ion is driven below the saturation limit$I/I_{sat}\approx 0.5$. }
\label{scheme}
\end{figure}

\begin{figure}
\centering
\includegraphics[width=1.0\textwidth]{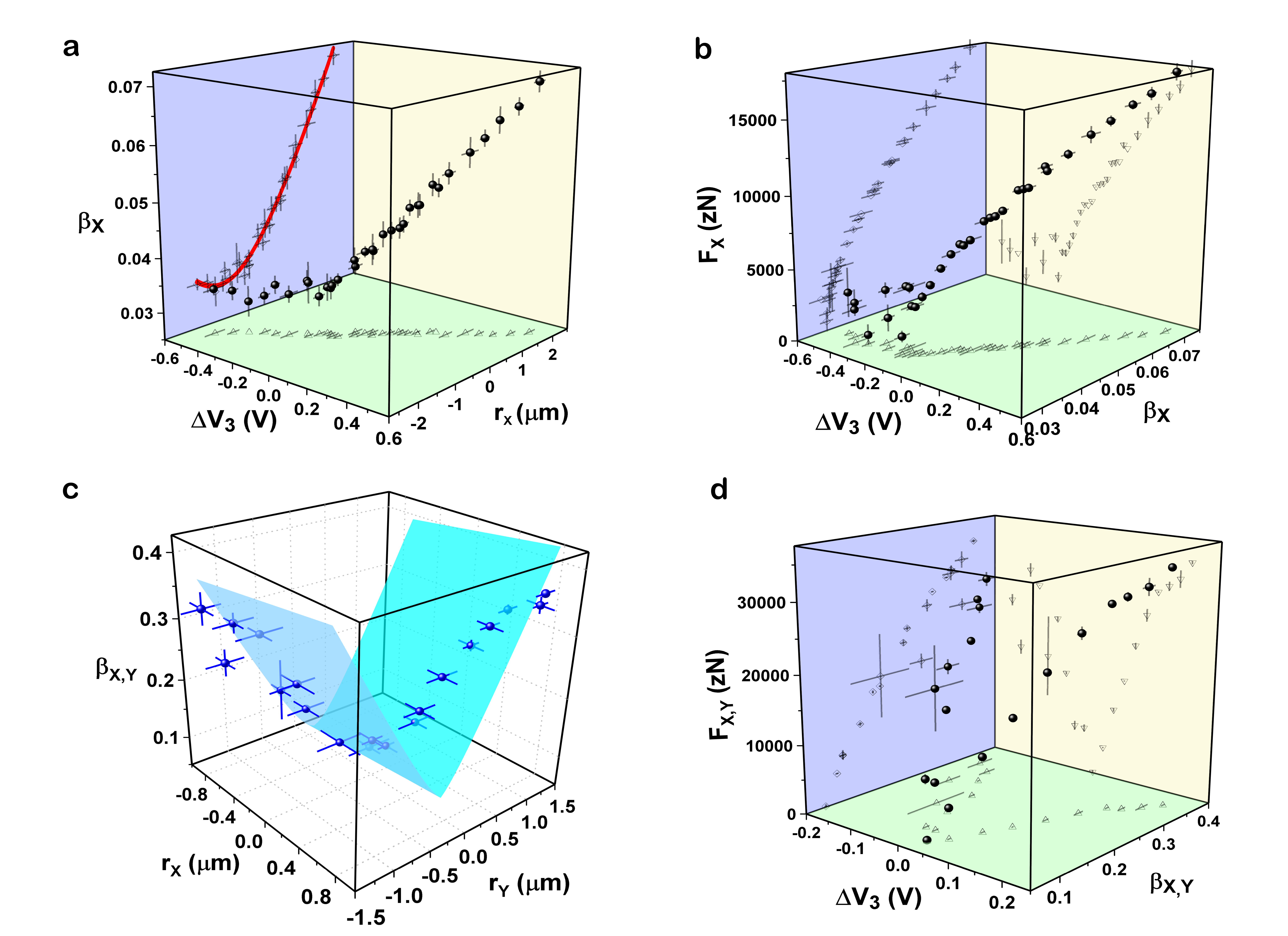}
\caption{The measurements of electric forces in geometry $I$. (a) The calibration of $\beta_X$ to the ion's axial
displacement $r_X$ at various $\Delta V_3$. The black spheres with error bars are experimental data from independent RF-photon correlation measurements and imaging measurements, while triangles (diamonds) are projections to the $\Delta V_{3}-r_{X}$ ($r_{X}-\beta_{X}$) plane, representing the dependence of the ion's axial displacement on the applied voltage (the correspondence of the micromotion index in trap axis to the ion's axial displacement). The red line on the $r_{X}-\beta_{X}$ plane is the theoretical fit to Eq.\ref{bx}. (b) The dependence of axial electric force $F_X$ on the applied voltage and micromotion index $\beta_X$. The triangles, inverted triangles and diamonds represent the dependence of the micromotion index $\beta_X$ on the applied voltage, axial force $F_X$ on the applied voltage and on the micromotion index $\beta_X$, respectively. (c) The calibration of $\beta_{X,Y}$ to the ion's displacement $r_X$ and $r_Y$ at various applied voltages. The blue spheres with error bars are experimental data, and the cyan surface is the theoretical fit to Eq.\ref{by}. (d) The dependence of electric force $F_{X,Y}$ on the applied voltage and micromotion index $\beta_{X,Y}$. The triangles, inverted triangles and diamonds represent the dependence of the micromotion index $\beta_{X,Y}$ on the applied voltage, force $F_{X,Y}$ on the applied voltage and on the micromotion index $\beta_{X,Y}$, respectively.}
 \label{v3}
\end{figure}

\emph{Near-saddle point case.} The initial ion's position is deliberately offset from the saddle
point, allowing for the measurement of the force up to a hundred atto-Newtons. We first carried out the force measurement in geometry $I$, where the voltage $V_3$ on electrode DC3 was varied in small steps. The micromotion indices, the trapping frequencies, and the ion's images were measured for each $V_3$, as shown in Figure~\ref{v3}. In Figure~\ref{v3}(a), we calibrated  $\beta_X$ to the ion's axial displacement
$r_X$ based on the theoretical relation
\begin{equation} \label{bx}
\beta_{X}=\sqrt{\beta^{2}_{X,eff}+\beta_{0}^2}=\sqrt{(\pi q_{X}(r_{X}-r_{X,0})/\lambda)^{2}+\beta^{2}_{0}},
\end{equation}
From this calibration, we determined non-zero $q_X = 0.00234(5)$ and $\beta_0 = 0.0318(3)$, which
were attributed to the non-vanishing RF field along the trap axis, and $r_{X,0} = -1.35(5) \mu m$
was the initial displacement from the RF node. It was evident that all $\beta_X$ could be well
calibrated by $r_X$ according to Eq.~\ref{bx}. It is noted that contrary to $q_y$ and $q_z$, which
can be deduced from the measurements of radial secular frequencies $\omega_y$ and $\omega_z$ with
varying RF voltages, $q_X$ cannot be obtained from trap frequency measurements, because the RF
potential is never dominant in this direction. By fitting the uncertainties of $q_X$, $\beta_0$ and
$r_{X,0}$ with this calibration, we estimate that the corrections for the force component $F_X$
along the axial direction of the linear trap is less than $3.4\%$. In Figure \ref{v3}(b), by
varying voltage on electrode DC3, the electric forces were attained by following relation modified
from Eq.\ref{fi}:
\begin{equation} \label{fx}
	F_X = \frac{\lambda m \omega^{2}_{X} \sqrt{\beta^{2}_{X}-\beta^{2}_{0}}}{\pi q_X}.
\end{equation}
Here the trapping frequency $\omega_X$ was measured independently. As is evident from Figure \ref{v3}(b), when the applied voltage difference
was beyond the range of $\Delta V_3 = (-0.5 V, -0.12 V)$, did $\beta_X$ and $F_X$ start to increase monotonically. In the range of
$5 < F_{X} < 20$ aN, it clearly demonstrated very good linear dependence on the applied voltage.

In Figure \ref{v3}(c), the calibration of $\beta_{X,Y}$ to the ion's displacement $r_X$ and $r_Y$ is
illustrated, which agrees well with the theoretical fit according to
\begin{equation} \label{by}
\beta_{X,Y}=\sqrt{(\pi q_{X}(r_{X}-r_{X,0})\sin \phi/\lambda + \pi q_{y,z}\cos \phi
(r_{Y}-r_{Y,0})/\lambda)^2 +\beta^{2}_{1}},
\end{equation}
whose derivation can be found in Methods Eq.\ref{bxy}. $\phi = 14^{\circ}$ is the angle between the direction of detection laser and $Y$ axis in the
lab coordinate, which was chosen for the minimization of the background scattering from the vacuum
viewport windows. $q_{y,z} = \sqrt{q_y q_z}=0.209$. $r_{Y,0} = -0.03(4) \mu m$ is the initial displacement from the RF node. $\beta_1 = 0.077(19)$ originates from
both the non-vanishing RF field along the trap axis and the residual phase shifts of the RF
potential on the opposing radial electrodes. Thus, based on the measurement of the micromotion
indices and the ion's position displacements, the electric forces were attained by
\begin{equation} \label{fy}
\begin{aligned}
F_{X,Y} &= m\omega^{2}_{Y}(r_{Y}-r_{Y,0})\\
&= \frac{\lambda m\omega^{2}_{Y}}{\pi q_{y,z}}\frac{\sqrt{\beta^2_{X,Y}-\beta^2_{1}}-\sqrt{\beta^2_{X}-\beta^2_{0}}\sin \phi}{\cos \phi},
\end{aligned}
\end{equation}
where $\omega_{Y}=\sqrt{\omega^2_{y}\cos ^2 \theta + \omega^2_{z}\sin^2 \theta}$ is the oscillation
frequency along the $Y$ axis, as shown in Figure \ref{v3}(d). Compared to the axial force $F_X$,
the radial force $F_{X,Y}$ is more sensitive and illustrates a nearly symmetric dependence on $\Delta
V_3$ as what we expected, despite of relatively large systematic uncertainty. The minimal detected
force is $F_X = 653.98 \pm 743.16$ zN and $F_{X,Y} =119.69 \pm 331.91 $ zN  in 30 s measurement time,
corresponding to a measurement sensitivity about 3582.00 $\textrm{zN}/\sqrt{\textrm{Hz}}$ and 655.57 $\textrm{zN}/\sqrt{\textrm{Hz}}$,
respectively~\cite{degen2017quantum}.

\begin{figure}
\centering\includegraphics[width=0.8\textwidth]{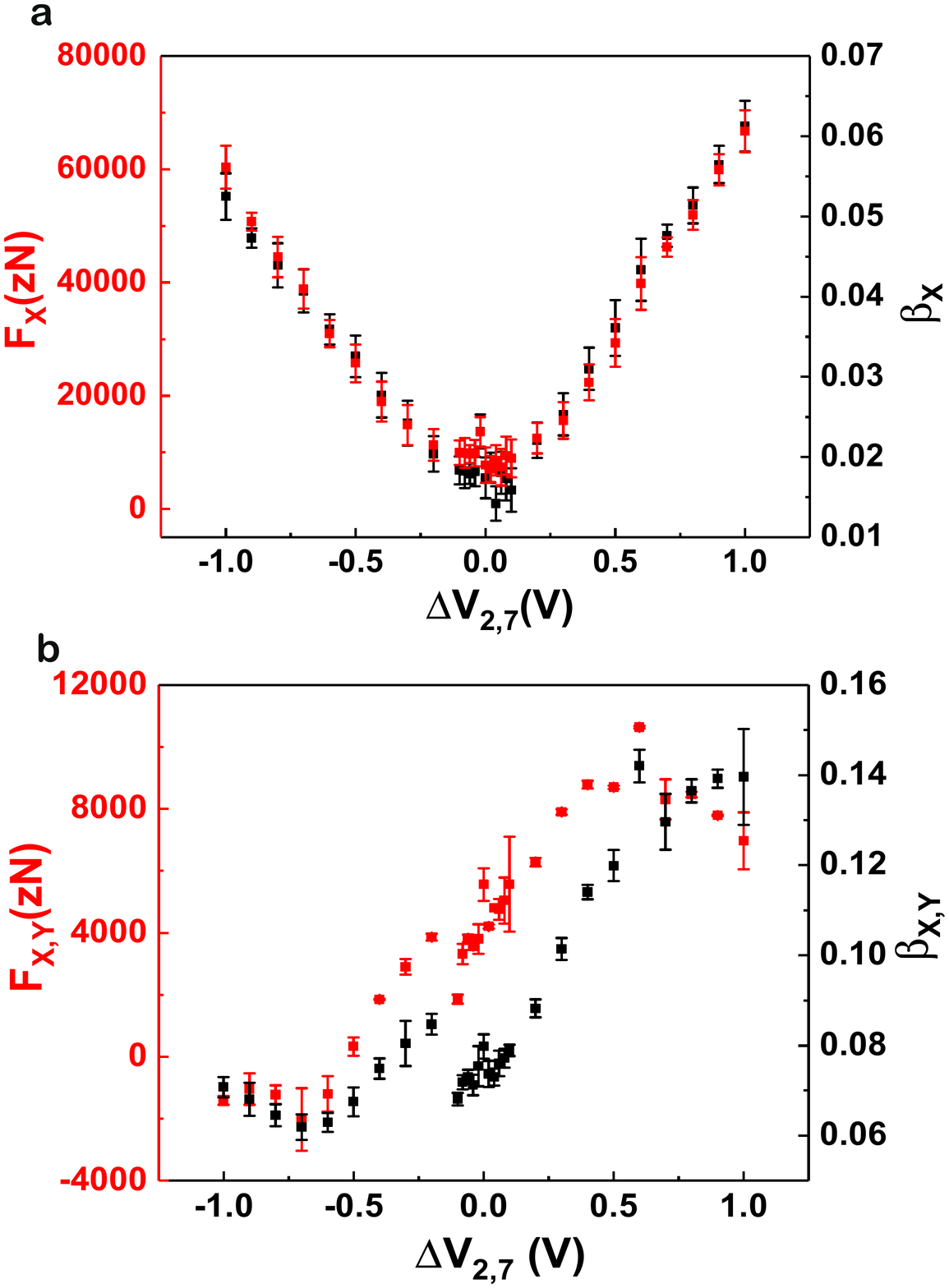}
\caption{The measurements of electric forces in geometry $II$. (a) The dependence of the micromotion index $\beta_X$ and force component $F_X$ on the $\Delta V_{2,7}$. (b) The dependence
of the micromotion index $\beta_{X,Y}$ and force component $F_{X,Y}$ on the $\Delta V_{2,7}$.}
\label{v2}
\end{figure}

The forces in geometry $II$, generated by varying the voltage on a pair of axial-symmetric
electrodes DC2 and DC7, were also measured. Figure~\ref{v2} shows the dependence of the axial force
$F_X$ and the radial force $F_{X,Y }$ on $\Delta V_{2,7}$. In Figure~\ref{v2}(a), $F_X$ varies nearly symmetric with $\Delta V_{2,7}$. $F_{X,Y }$ is more sensitive to $\Delta V_{2,7}$ compared with $F_X$. The minimal detected force is $F_X = 951.13 \pm
3030.95 $ zN and $F_{X,Y }= 329.39 \pm 295.42 $ zN in 30 s measurement time, corresponding to a
measurement sensitivity about 5209.55 $\textrm{zN}/\sqrt{\textrm{Hz}}$ and 1804.14 $\textrm{zN}/\sqrt{\textrm{Hz}}$, respectively.

\begin{figure}
\centering
\includegraphics[width=1.0\textwidth]{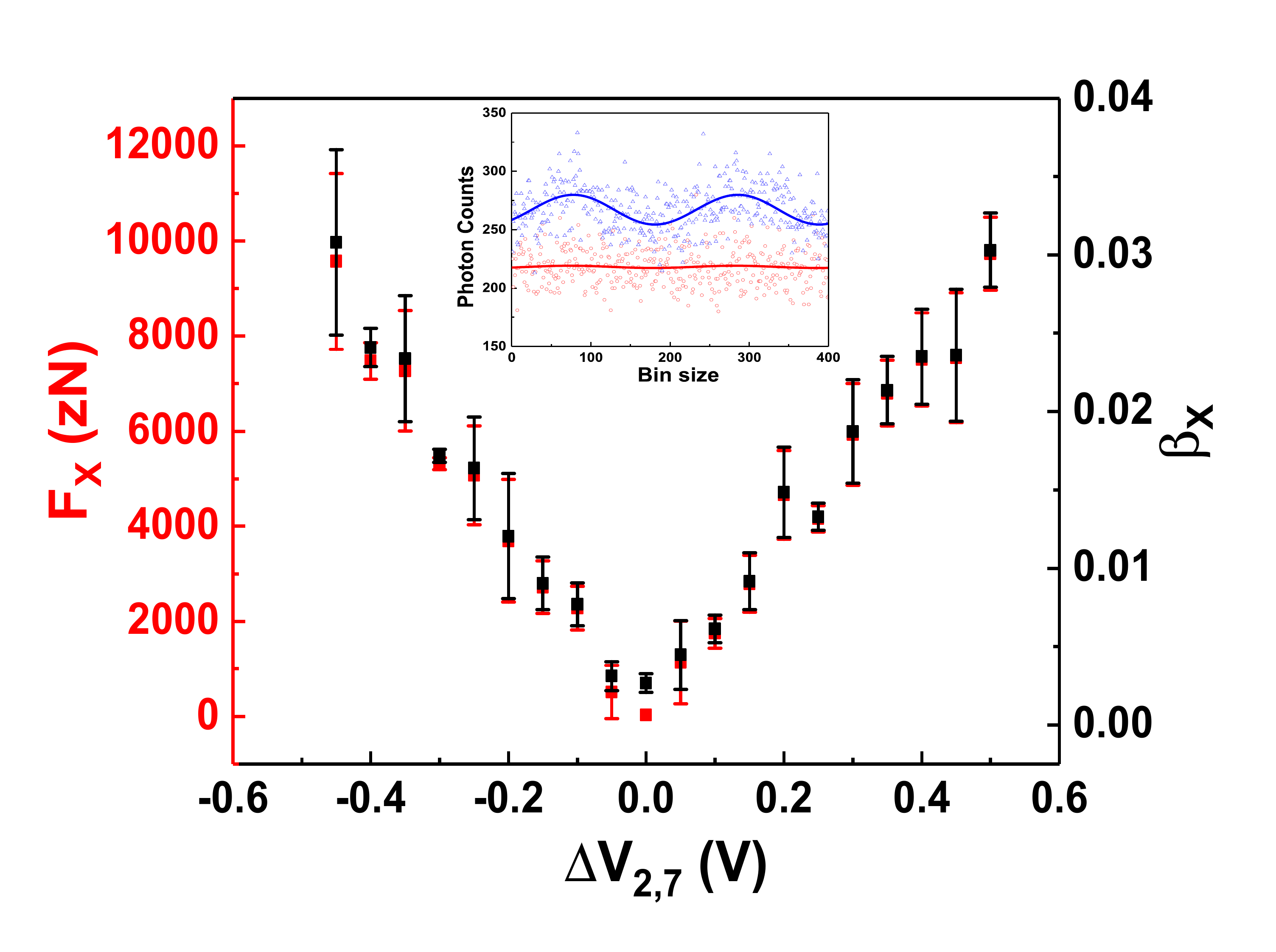}
\caption{Electric force Detection at the saddle point. The dependence of $\beta_x$ on $\Delta
V_{2,7}$. The micromotion is minimized at the saddle point with the micromotion indices of
$\beta_X=0.0027(6)$ and $\beta_Y=0.0308(28)$, and $q_{y}=0.1827$ and $q_{z}=0.1840$  obtained
from the trapping frequency dependence on the applied RF voltage. $V_2$ and $V_7$ were varied
simultaneously in small voltage step $\Delta V=50$ mV, and their relation was $V_{7}=V_{2}+0.25V$.
In the inset, the red circle (blue triangle)  and solid line correspond to the histogram measured at
$V_{2}=11.75V$ ($V_{2}=12.25V$) and its fitting, respectively. The corresponding micromotion index
is $\beta_X=0.0030$ ($\beta_X=0.0306$).}
\label{fig3}
\end{figure}

\emph{At saddle point case.} We performed the force measurements exactly at the saddle point. In
order to investigate the best force sensitivity, the ion is close to the saddle point as possible
as we could by by the applying machine learning approach~\cite{liu2021minimization}. By applying
voltages on DC2 and DC7, we obtain a linear dependence of $\beta_X$ on the applied voltage $V_2$
and $V_7$, shown in Figure \ref{fig3}, while $\beta_Y \approx 0$ for all the voltage sets,
indicating that the ion moves along the trap axis $X$. The applied force is then deduced according
to Eq.~\ref{fx}. The inset in Figure~\ref{fig3} shows the histogram measured at $V_{2}=11.75$ V
(red circles) and $V_{2}=12.25$ V (blue triangles) as well as their theoretical fits
(solid lines). The corresponding micromotion index are $\beta_X=0.0030$ and $\beta_X=0.0306$. The
minimal detected force is $F_X = 514.37 \pm 560.78 $ zN  in 30 s measurement time, corresponding to
a measurement sensitivity about 2817.32 $\textrm{zN}/\sqrt{\textrm{Hz}}$, which is about 2 fold better than near saddle-point case.

We have demonstrated a proof-of-principle experiment for measuring extremely weak electric forces
using a mechano-optical transducer with a single trapped ion. Comparing with other sensing
techniques with trapped ions, our method of mechano-optical transduction can measure a unknown
force without having the knowledge of their frequency spectrum, thus distinguishing itself for 3D
sensing of force anomaly. For this application, electric forces in two different geometries have
been accurately measured by alternating the detection laser between $X$ and $Y$ axes, illustrating
an excellent 3D force sensor given full optical access in three dimensions. Furthermore, electric
forces down to $F_X = 514.37 \pm 560.78 \textrm{zN}$ and $F_{X,Y}= 119.69 \pm 331.91 \textrm{zN}$ have been acquired in
30 s data acquisition time with a sensitivity of 2817.32 $\textrm{zN}/\sqrt{\textrm{Hz}}$ and 655.57$\textrm{zN}/\sqrt{\textrm{Hz}}$
in $X$ and $Y$ axis, respectively. The experimental uncertainty includes statistical fluctuations
in averaging over the ion fluorescence, the uncertainty in the fitted micromotion index, the
uncertainty in the calibration of the applied electric field due to relatively large coma
abberations in the ion image, the uncertainty in the measured trapping frequencies, and the
uncertainty in the trap stability q-parameter.

Further improvements are possible for a better sensitivity. First, detection efficiency can be
enhanced by realistic modifications and update of current experimental setup, for instance, larger
solid angle for fluorescence light collection enabled by a binary phase Fresnel lens with $NA =
0.64$~\cite{blums2018single} or by integration with a glass vacuum cell~\cite{he2021ion}, using a
superconducting single photon detectors with $\approx 80\%$ detection
efficiency~\cite{crain2019high,todaro2021state} compared to current photon multiplier tube with a
nominal 32$\%$ at 370 nm. Adapting both improvements would result in a factor of $\times4$
enhancement in detection efficiency, providing up to 2-fold increase in the force detection
sensitivity. According to Eq.~\ref{fi} and the pseudopotential approximation
\cite{paul1990electromagnetic},
\begin{equation}
\begin{aligned}
F_{i} &\approx \frac{\lambda m\beta_i \Omega^2 }{4 \pi }(\frac{a_i}{q_i}+\frac{q_i}{2})\\
&\ge \frac{\lambda mq_i\beta_i \Omega^2 }{4 \pi },
\end{aligned}
\end{equation}
where $\Omega$ is the RF driving frequency. The minimal detectable electric force can be achieved
for $q_{i}=\sqrt{2a_i}$. Therefore, additional optimization of the trapped-ion setup, such as
trapping a different species with smaller mass and reducing the driving RF frequency could allow us
to further enhance the sensitivity by more than four orders of magnitude. As an exanple, with
single trapped $^9 Be^+$~\cite{meekhof1996generation}, the RF frequency $\Omega = 2\pi \times$ 1.4
MHz, $q_{i}=0.017$, $\beta_{i}=0.001$ for a segmented linear RF trap~\cite{schmidt2020mass}, it is
possible to discriminate electric force of well below 1 $yN$ (1 $yN=10^{-24} N$) with a sensitivity
$\approx$ 2 $yN/\sqrt{Hz}$. Such a single-atom level mechano-optical transducer would provide
opportunities to search for the fifth-force beyond the standard model~\cite{safronova2018search},
including exotic spin-dependent force
\cite{dobrescu2006spin,kotler2014measurement,kotler2015constraints,chen2016stronger,rong2018searching,almasi2020new}
and other types of the new forces~\cite{ding2020constraints,jiao2021experimental}.

\textbf{Acknowledgements} This work is supported by the Key-Area Research and Development Program
of Guang Dong Province under Grant No.2019B030330001, the National Natural Science Foundation of
China under Grant No.11774436, No.11974434 and No. 12074439. Le Luo received supports from
Guangdong Province Youth Talent Program under Grant No.2017GC010656, Sun Yat-sen University Core
Technology Development Fund. Yang Liu acknowledges the support from Natural Science Foundation of
Guangdong Province under Grant 2020A1515011159, and Science and Technology Program of Guangzhou, China 202102080380.

\bibliography{weak_force}
\newpage

\textbf{Methods} In a typical Paul trap, when the ion is disturbed by an additional static
electric field $E_{dc}$, its equation of motion can be written as \cite{berkeland1998minimization}:
\begin{equation} \label{field}
\ddot{u}_{i}+[a_{i}+2q_{i}cos(\Omega t)]\frac{\Omega^{2}}{4}u_{i}=\frac{qE_{i}}{m}.
\end{equation}
For $|q_{i}|\ll 1$ and $|a_{i}|\ll 1$, the solution to lowest order in $q_i$ and $a_i$ is
\begin{equation} \label{solution}
u_{i}(t) \cong [r_{i}+s_{i}cos(\omega_{i}t + \phi_{si})][1+\frac{q_{i}}{2}cos(\Omega t +
\phi_{i})],
\end{equation}
where $r_{i}\cong \frac{4eE_{i}}{m(a_{i}+\frac{q_{i}^{2}}{2})\Omega^{2}} \cong
\frac{eE_{i}}{m\omega_{i}^2}$ is a displacement of the equilibrium position from the rf node for a
trapped ion with charge $e$ and mass $m$. $s_i$ is the secular motion amplitude with the frequency
of $\omega_{i} = \frac{\Omega}{2}\sqrt{a_{i}+\frac{q_{i}^{2}}{2}}$. $\phi_{si}$ is the phase of
secular motion determined by the initial conditions of the ion position and velocity. $\phi_{i}$ is
the phase of the micromotion.

This solution describes that for a trapped ion near the saddle point of the RF trapping potential, a weak electric force $q\vec{E}_{i}$ would not only shift the ion's equilibrium position
by $r_i$, but also induces an excess micromotion modulated by the RF frequency of $\Omega$ with an amplitude of
$\frac{1}{2}r_{i}q_{i}$ along $\hat{u}_i$ at position $r_i$. With the assumption of zero phase
difference between two ac electrodes $\phi_{ac}=0$, the micromotion index is described by
$\beta_{i} =\frac{1}{2}k_{i}q_{i}|r_{i}|$, where $k_i$ is the component of wave vector on axis $i$.
Since $q_i$ is proportional $V_{RF}/\Omega^2$ and is dependent on the trap's geometric
configuration, it would not change by varying the DC voltages, therefore the micromotion index is
proportional to the ion's displacement $\beta_{i}\propto r_i$. Because of the coefficient
$kq_{i}/2=\pi q_{i}/\lambda \gg1$ for a typical Paul trap (in our case, $q_{x}\sim 4\times10^{-4}$
and $q_{y,z}\sim 0.2$) and ultraviolet (or visible) lasers, it would bring a large amplification
effect on measuring a weak force. Thus, an extremely weak electric force which cannot be detected
by displacement measurement from imaging method, can be precisely measured through measurement of
$\beta_i$ using RF-photon correlation technique~\cite{keller2015precise}.

In our experiment, weak unknown force is generated by applying a
small electric DC voltage to the DC electrodes of the trap. Then the excess micromotion generates a
fluorescence signal modulated by the RF frequency. The fluorescence is detected by the
photomultiplier tube (PMT) and is successively analyzed by the time-to-digital converter (TDC). The
experimental schematics, detailed experimental sequences and demodulated ion's fluorescence are shown in Figure \ref{scheme}. The micromotion index $\beta_i$ is inferred by fitting the recorded fluorescence histogram to the model of damped harmonic oscillator \cite{keller2015precise,liu2021minimization}. The force related to the micromotion is determined by
\begin{equation} \label{fi}
	F_i = \frac{\lambda m \omega^{2}_{i} \beta_i}{\pi q_i}\propto \Delta V,
\end{equation}
since $\beta_i = \frac{1}{2}k_i q_i r_i = \pi q_i e E_i / \lambda m \omega^{2}_i$. Therefore, a three-dimensional force can be
precisely determined from micromotion indices, which can be measured simply by alternating the
detection laser between three axes, i.e. the three principal axes of the trap. Such a method have a
intrinsical lock-in function, where the modulated and demodulated signal are mechanical and
optical, respectively.

Consequently, this system constitutes a mechano-optical transducer at a single
atom level. Such a transducer not only has extremely high spatial resolution, but also has the
advantages over standard lock-in technique, because it continuously scans the phase between
modulated and demodulated signal, thus enabling the detection of force of unknown frequency
spectrum. For developing a force sensor, there usually has to be a compromise between detection
sensitivity and measurement range. In our system, the best force sensitivity can be achieved when
the ion's unperturbed position exactly at the saddle point, while the larger force range can be
obtained when the ion's unperturbed position have a limited offset to the saddle point.

In this method, since the fluorescence is modulated by the RF frequency and its second harmonics~\cite{berkeland1998minimization, keller2015precise}, any noise which is hardly in
phase with the RF frequency, and averaging over $N \gg 1$ times, would greatly enhance the contrast
of the desired signal relative to the background noise by a factor of $N^{1/2}$, similar to the
phase-coherent Doppler velocimetry \cite{biercuk2010ultrasensitive}.

Practically, in order to realize 3D force sensing, the force measurement were carried out in two different force geometries by varying the voltages of the different DC electrodes. In the force detection for both geometries, the axial and radial
forces were measured by alternately switching the detection laser between axis $X$ (the trap axis)
and $Y$ in lab coordinate. In geometry $I$, the forces are applied by only varying the voltage on electrode DC3, while in geometry $II$, the forces are applied
by only varying the voltage on electrodes DC2 and DC7. The ion's displacements are also measured by recording the ion's
images at various voltage $V_i$ for calibration purposes. From these measurements, the dependence of the
$\beta_i$ on the applied voltage $V_i$, as well as the dependence of the ion's displacement $r_i$
on $V_i$, which provides a direct calibration of the relation between the micromotion index
$\beta_i$ and the ion's displacement $r_i$ using two independent methods, are obtained.

The calibration of $\beta_{X,Y}$ to the ion's displacement $r_X$ and $r_Y$ is given by
\begin{equation}
\label{bxy}
\begin{split}
\beta_{X,Y} & = \sqrt {{{({\beta _{X,eff}}\sin \phi  + {\beta _{y,eff}}\cos \phi \cos \theta  + {\beta _{z,eff}}\cos \phi \sin \theta )}^2} + \beta _1^2} \\
& = \sqrt {\begin{array}{l}
(\pi {q_X}({r_X} - {r_{X,0}})\sin \phi /\lambda  + \pi {q_y}({r_y} - {r_{y,0}})\cos \phi \cos \theta /\lambda \\
 + \pi {q_z}({r_z} - {r_{z,0}})\cos \phi \sin \theta /\lambda {)^2} + \beta _1^2
\end{array}} \\
& \simeq\sqrt {\begin{array}{l}
(\pi {q_X}({r_X} - {r_{X,0}})\sin \phi /\lambda  + \pi {q_{y,z}}\cos \phi [({r_y} - {r_{y,0}})\cos \theta \\
 + ({r_z} - {r_{z,0}})\sin \theta ]/\lambda {)^2} + \beta _1^2
\end{array}}
\\& = \sqrt{(\pi q_{X}(r_{X}-r_{X,0})\sin \phi/\lambda + \pi q_{y,z}\cos \phi
(r_{Y}-r_{Y,0})/\lambda)^2 +\beta^{2}_{1}}
\end{split}
\end{equation}
where $\theta \approx 53^{\circ}$ is the angle between $Y$ axis and principal axis $y$ in the radial
plane. Its variation with the applied voltage was much smaller than one degree and thus was ignored
in this analysis. $r_{y,0}$ and $r_{z,0}$ is the initial displacement in principal axes $y$ and
$z$, respectively. $q_y$ and $q_z$ are linearly proportional to the RF voltage theoretically, and
$q_{y}\approx 0.203(2)$ and $q_{z}\approx 0.215(3)$ were obtained from measurements of radial
trapping frequency as a function of RF voltage.

\textbf{Experimental Setup} Our force detection is performed on a single $^{171}$Yb$^+$ ion confined
in a linear Paul trap similar to Ref [\citenum{liu2021minimization}]. It consists of four gold-plated ceramic blade electrodes, where two
opposite blades labelled as RF1 and RF2 are driven with an RF potential, creating the transverse
(y-z) quadrupole confinement potential and other two opposite blades each have five segments
serving to confine the ion along the x-axis. The RF voltages are fed through a pair of home-made
helical resonator (quality factor Q$\approx$ 300) and a directional coupler (Mini Circuits
ZEDC-15-2B) and a RF amplifier (Mini Circuits LZY-22+) to the RF source (Rigol DSG821). Each
electrode is connected to a programmable precision DC voltage power supply (BS1-16-14,
Stahl-electronics). The gap between the electrodes is 470 $\mu m$  and 220 $\mu m$  in Y and Z
direction, respectively. The distance from the trap center to the electrodes R $\approx$ 259 $\mu
m$ . The cooling laser is 369.5 nm from a frequency-doubled Ti: Saphire laser (M Squared Lasers
Ltd, ECD-F) with its fundamental frequency is stabilized using acousto-optic modulation transfer
spectroscopy of Iodine, and the repumping lasers are 935 nm and 638 nm lasers (Toptica, DLC DL
PRO).

The ion images in x-y lab coordinate plane are recorded with an electron-magnified charge-coupled
device (Andor, ixon-ultra-897 EMCCD). In our imaging optics, the isotropic fluorescence from the
ion at $\lambda $=369.5 nm is transmitted through a vacuum viewport and collected by an objective
lens of numerical aperture NA = 0.397 with $\times$ 10 magnification. The intermediate image passes
through a pinhole for spatial filtering and subsequently magnified by another lens with $\times$ 50
magnification in total.

The ion fluorescence passes the same imaging ion optics, and the
temporal profile of ion fluorescence due to the applied drive is detected by recording
scattered-photon arrival times using a photomultiplier tube (H10682-210, Hamamatsu) relative to a
homemade edge-triggered D-type flip-flop electronics. Photon arrival times are then determined over $N$ iterations of the
experiment through demodulation using a START-STOP time-to-digital converter (HRM-TDC, SensL).

For the force detection near saddle-point, the initial voltages on electrodes
DC1, DC2, $\cdots$ DC12 were 9.69 V, 9.88 V, 1.867 V, 10.05 V, 10.05 V, 9.69 V, 9.88 V, 2.0 V,
10.05 V, 10.05 V, 0.186 V and 0.01 V, respectively. The amplitude of RF voltage was 206.48 V with
the frequency of $2\pi \times 21.250$ MHz. The corresponding trapping frequencies were
$\omega_{x}\sim 2\pi \times 721.66$ kHz, $\omega_{y}\sim 2\pi \times 1.184$ MHz and $\omega_{z}\sim
2\pi \times 1.602$ MHz, respectively.

While for the force detection at saddle-point, the applied voltages were $(V_{1}, V_{2}...V_{12},V_{RF})$=(11.75 V, 11.75 V, 1.754 V,
9.35 V, 9.35 V, 12 V, 12 V, 1.9 V, 9.6 V, 9.6 V, -0.079 V, -0.22 V, 227.84 V) at 0.45W RF power
and $\Omega=2\pi \times 22.625$ MHz RF frequency. The corresponding trapping frequencies were
$(\omega_x, \omega_y, \omega_z) = (2\pi \times 0.748, 2\pi \times 1.382, 2\pi \times 1.655)$ MHz.

\end{document}